\documentclass[%
 reprint,
 amsmath,amssymb,
 aps,
]{revtex4-2}

\usepackage{graphicx}% Include figure files
\usepackage{dcolumn}% Align table columns on decimal point
\usepackage{bm}% bold math
\usepackage{enumitem}

\makeatletter
\newcommand*{\clearpageandresetcolumns}{
  \par
  \close@column@grid
  \clearpage
  \twocolumngrid
  \par
  \vspace*{0pt}
}
\makeatother

\setlist{leftmargin=*}

\begin{document}

\preprint{APS/123-QED}

\title{\textbf{Wakefield-Dressed Relativistic Vortex Electrons in Plasma Accelerators}}% Force line breaks with \\
% \title{Wakefield-Dressed Vortex Eigenstates of Relativistic Electrons}% Force line breaks with \\

\author{Zhigang Bu}
\author{Lingang Zhang}
\author{Liangliang Ji}
\email{jill@siom.ac.cn}
\affiliation{%
State Key Laboratory of Ultra-intense Laser Science and Technology, Shanghai Institute of Optics and Fine Mechanics (SIOM), Chinese Academy of Sciences (CAS), Shanghai 201800, China
% \textbackslash\textbackslash
}%

\date{\today}% It is always \today, today,
             %  but any date may be explicitly specified

\begin{abstract}
Plasma wakefield acceleration is usually regarded as a classical mechanism for producing high-energy charged-particle beams. Here we show that an axisymmetric plasma wakefield can also act as a moving quantum structure that supports relativistic vortex electron states. Starting from the Dirac equation, we derive the electron spinor eigenstates with definite total angular momentum in an ideal bubble-regime wakefield. The transverse focusing field confines and quantizes the electron transverse motion into Laguerre-Gaussian vortex modes, while the longitudinal electric field accelerates the electron without destroying the symmetry-protected angular momenta. We further analyze the localized and off-axis vortex electron wave-packets, and non-ideal wakefield perturbations, and identify the conditions for preserving electron-vortex-state purity. These results suggest plasma wakefield as a route toward high-energy vortex electrons and extend plasma-based acceleration from classical beam dynamics to quantum-state control of relativistic particles.
\end{abstract}

%\keywords{Suggested keywords}%Use showkeys class option if keyword
                              %display desired
\maketitle

%\tableofcontents

\textit{Introduction}---Vortex states are quantum states with helical phases and well-defined orbital angular momentum (OAM) at the single-particle level. Since the discovery of optical vortex beams \cite{r01,r02,r03,r04,r05}, vortex states have been extended to electrons \cite{r06,r07,r08}, neutrons \cite{r09,r10}, and atoms \cite{r11,r12}, enabling applications in particle manipulation \cite{r13}, quantum information \cite{r14,r15,r16}, imaging techniques \cite{r17}, and materials science \cite{r18,r19}. Electron vortex beams are especially attractive because their OAM and associated magnetic moment provide additional degrees of freedom for manipulating the microscopic quantum systems. To date, electron vortex states have been demonstrated at low and moderate energies, typically up to hundreds of keV \cite{r06,r07,r08}.  Extending them to the high-energy regime remains a significant challenge. 

High-energy vortex electrons combine the vortex phase structure with short de Broglie wavelength, with possible relevance to high-energy physics\cite{r20}, nuclear physics and astrophysics \cite{r21,r22,r23,r24,r25}. Since direct generation become increasingly difficult as the de Broglie wavelength decreases, a natural route is therefore to prepare vortex electrons at lower energies and subsequently accelerate them. This strategy raises a fundamental question: can an accelerating field preserve, or even support, a well-defined quantum vortex state of a relativistic electron? 

This question has recently been considered in conventional accelerator settings. Quasi-classical studies have shown that electron OAM precession in accelerator fields may induce resonance that disrupts the vortex structure \cite{r26}. Linacs, where such complications are ignorable, have therefore been proposed as a possible route toward relativistic vortex electrons. More generally, existing theoretical treatments of vortex-electron dynamics in external fields are often based on nonrelativistic theory \cite{r27}, quasiclassical \cite{r26} or weak-field approximations \cite{r28} for the precessions of expectation values of OAM and spin, or scalar models that neglect the spinor structure of the relativistic electron states \cite{r29}. A relativistic description of detailed spinor vortex states in strong, structured accelerating fields is still needed.

In this letter, we find that plasma wakefield provide a natural setting for this problem. Driven by an intense laser pulse \cite{r30,r31,r32,r33} or charged-particle beam \cite{r34,r35,r36}, a plasma wakefield accelerator (PWFA) offers accelerating gradients far beyond those of conventional radio-frequency accelerators \cite{r37}. In the nonlinear bubble regime, the wakefield combines a strong longitudinal accelerating field with transverse focusing field near the propagation axis \cite{r38,r39,r40}. An approximately axisymmetric plasma wake can therefore simultaneously accelerate electrons, confine their transverse modes, and protect angular momentum through rotational symmetry.

Starting from the Dirac equation, we demonstrate that an axisymmetric plasma wakefield admits relativistic spinor vortex eigenstates with definite total angular momentum (TAM). Based on the spinor description of vortex states in crossed-field firstly developed in this work, we find that the transverse focusing field quantizes the electron motion into Laguerre-Gaussian (LG) vortex modes, while the longitudinal electric field accelerates the vortex electron without breaking the symmetry-protected angular momentum. We also analyze how off-axis injected electron wave-packets and non-ideal wakefield perturbations lead to the broadening of angular momentum spectrum. These results establish plasma wakefields as moving quantum structures capable of supporting and accelerating relativistic vortex electrons, providing a foundation for extending PWFA physics from classical beam dynamics to quantum-state control.

\begin{figure}[t]
\centering
\includegraphics[width=8.6cm]{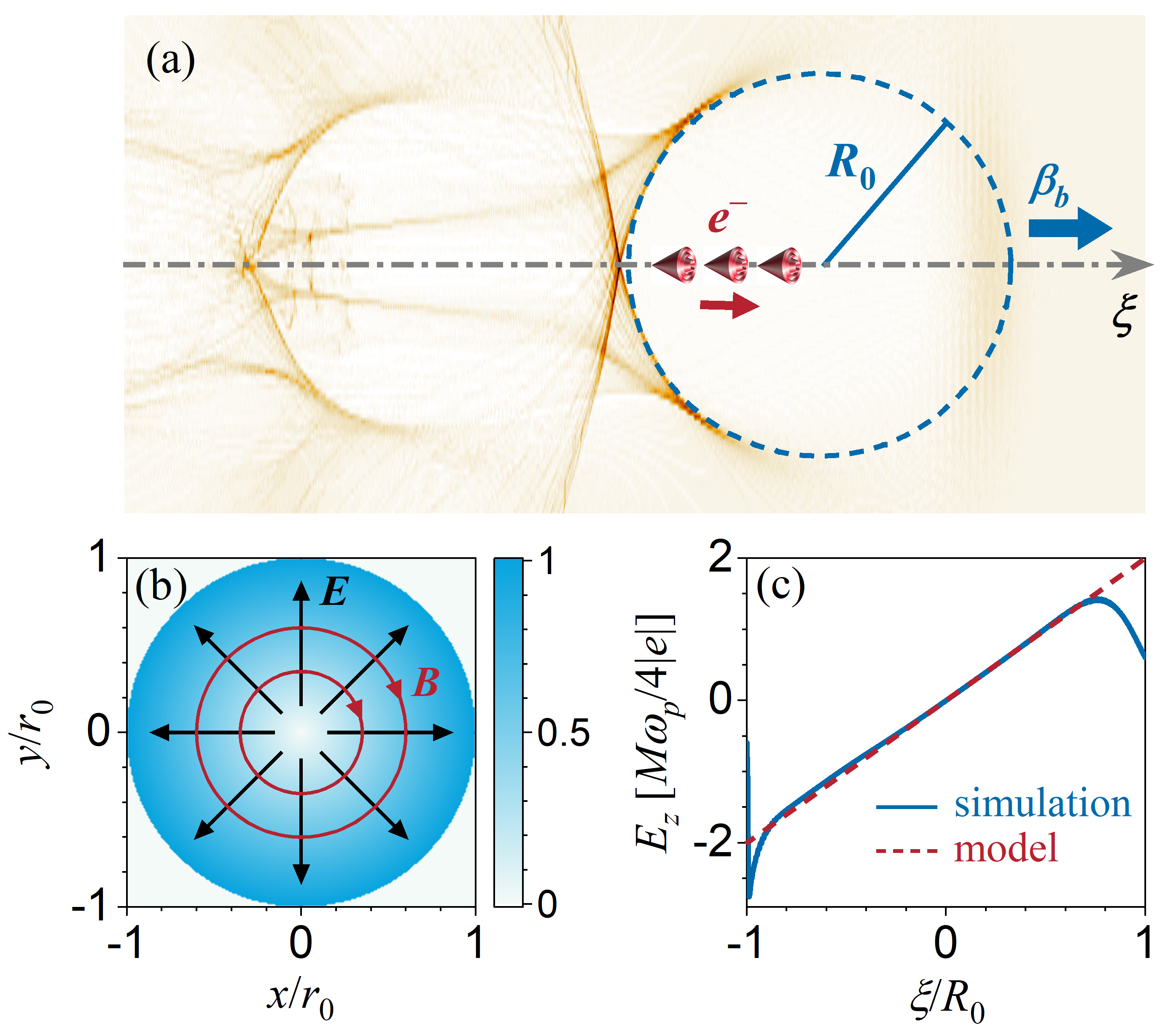}
\caption{Physical model. (a) Schematic layout of vortex electron acceleration in plasma wakefield. The orange background represents the electron density distribution of plasma bubble driven by an intense laser pulse. (b) Transverse electric and magnetic field distributions in the wake. (c) Longitudinal accelerating field on the axis.}
\label{fig:01}
\end{figure}

\textit{Physical model}---We model the nonlinear plasma wakefield in bubble regime by the electromagnetic field inside a spherical electron cavity of radius $R_{0}$ moving with velocity $\beta_{b}$ (relativistic factor $\gamma_{b}$) in a uniform plasma \cite{r39,r40}, see Fig.~\ref{fig:01}(a). The four-potential of this wakefield can be written, in the gauge of $A_{z}=-A_{0}$, as $A^{\mu}(t,\bm{x})=(A_{0}(t,\bm{x}),\bm{A}(t,\bm{x}))=\varphi(t,r,z)(-1,0,0,1)$, with $\varphi(t,r,z)=\varphi_{0}[r^{2}+(z-\beta_{b}t)^{2}]$ inside the cavity \cite{r39}. Here $r^2=x^2+y^2$, $\varphi_{0}=m_{e}\omega_{p}^{2}/8|e|$, $\omega_{p}=\sqrt{4\pi e^{2}n_{0}/m_{e}}$ is the plasma frequency, $n_{0}$ is the plasma density, and $m_{e}$ and $e$ are the electron mass and charge, respectively. The corresponding electric and magnetic fields are $\bm{E}(t,\bm{x})=2\varphi_{0}(x,y,(1+\beta_{b})(z-\beta_{b}t))$ and $\bm{B}(x,y)=2\varphi_{0}(y,-x,0)$, as shown in Fig.~\ref{fig:01}(b) and (c). These fields contain a longitudinal electric field for acceleration, and a radial electric field and an azimuthal magnetic field for focusing. The quadratic dependence of $\varphi$ on $r$ is essential: near the axis, the transverse force is linear in the displacement from the wake axis. At the quantum level, this focusing force acts as a harmonic-oscillator-like potential for confining the electron state.

The wave-function of an electron dressed in the wakefield is governed by the Dirac equation, 
\begin{equation}
\left[\gamma^{\mu}\left(i\partial_{\mu}-eA_{\mu}(t,\bm{x})\right)-m_{e}\right]\Psi(t,\bm{x})=0,
\label{eq:01}
\end{equation}
where $\gamma^{\mu}$ is the Dirac matrix. The potential of an ideal wakefield is axisymmetric and $\tau$-independent in co-moving variables. Hence the Dirac equation admits a vortex solution with conserved energy. The Dirac Hamiltonian commutes with the TAM projection, $[H_{0}, J_{z}]=0$, where $J_{z}=L_{z}+\Sigma_{z}/2$, $L_{z}=-i\partial_{\theta}$ is the projection of the OAM operator and $\Sigma_{z}$ is the projection of the spin operator $\bm{\Sigma}=\text{diag}(\bm{\sigma},\bm{\sigma})$. Thus, $H_{0}$ and $J_{z}$ share the common vortex eigenstates, and electron TAM is conserved in the wakefield. In the following, we construct the wakefield-dressed spinor eigenstates and show that they possess definite TAM and vortex phase structure.

\textit{TAM eigenstate in plasma wakefield}---Resolving Dirac equation in complex external fields is generally challenging. Let us seek a spinor solution in the following form,
\begin{alignat}{1}
&\Psi(t,\bm{x})\nonumber\\
=&\,N\left[\gamma^{0}\left(i\partial_{t}-eA_{0}\right)+\gamma^{0}\bm{\Sigma}\cdot\left(i\nabla+e\bm{A}\right)\left(\bm{\Sigma}\cdot\bm{n}\right)+m_{e}\right]\nonumber\\
&\,\times\bigg(\begin{array}{c}
\left[(1+\bm{\sigma}\cdot\bm{n})\Phi_{+}+(1-\bm{\sigma}\cdot\bm{n})\Phi_{-}\right]\chi^{s} \\
\left[(1+\bm{\sigma}\cdot\bm{n})\Phi_{+}-(1-\bm{\sigma}\cdot\bm{n})\Phi_{-}\right]\chi^{s}
\end{array}\bigg),
\label{eq:02}
\end{alignat}
where $\Phi_{\pm}(t,\bm{x})$ are scalar wave-functions, $\chi^{s}$ is the two-component spinor wave-function of free electron, $N$ is a normalization factor. Using solution \eqref{eq:02}, the Dirac equation can be reduced into two uncoupled scalar equations,
\begin{equation}
\big[(i\partial_{\mu}-eA_{\mu})^{2}-m_{e}^{2}-ie(\bm{E}\cdot\bm{n})\pm e(\bm{B}\cdot\bm{n})\big]\Phi_{\pm}(t,\bm{x})=0,
\label{eq:03}
\end{equation}
when the transverse components of the external fields satisfy the crossed-field condition, $\bm{B}_{\perp}=\bm{n}\times\bm{E}_{\perp}$, with the longitudinal unit vector $\bm{n}$.

\begin{figure}[t]
\centering
\includegraphics[width=8.6cm]{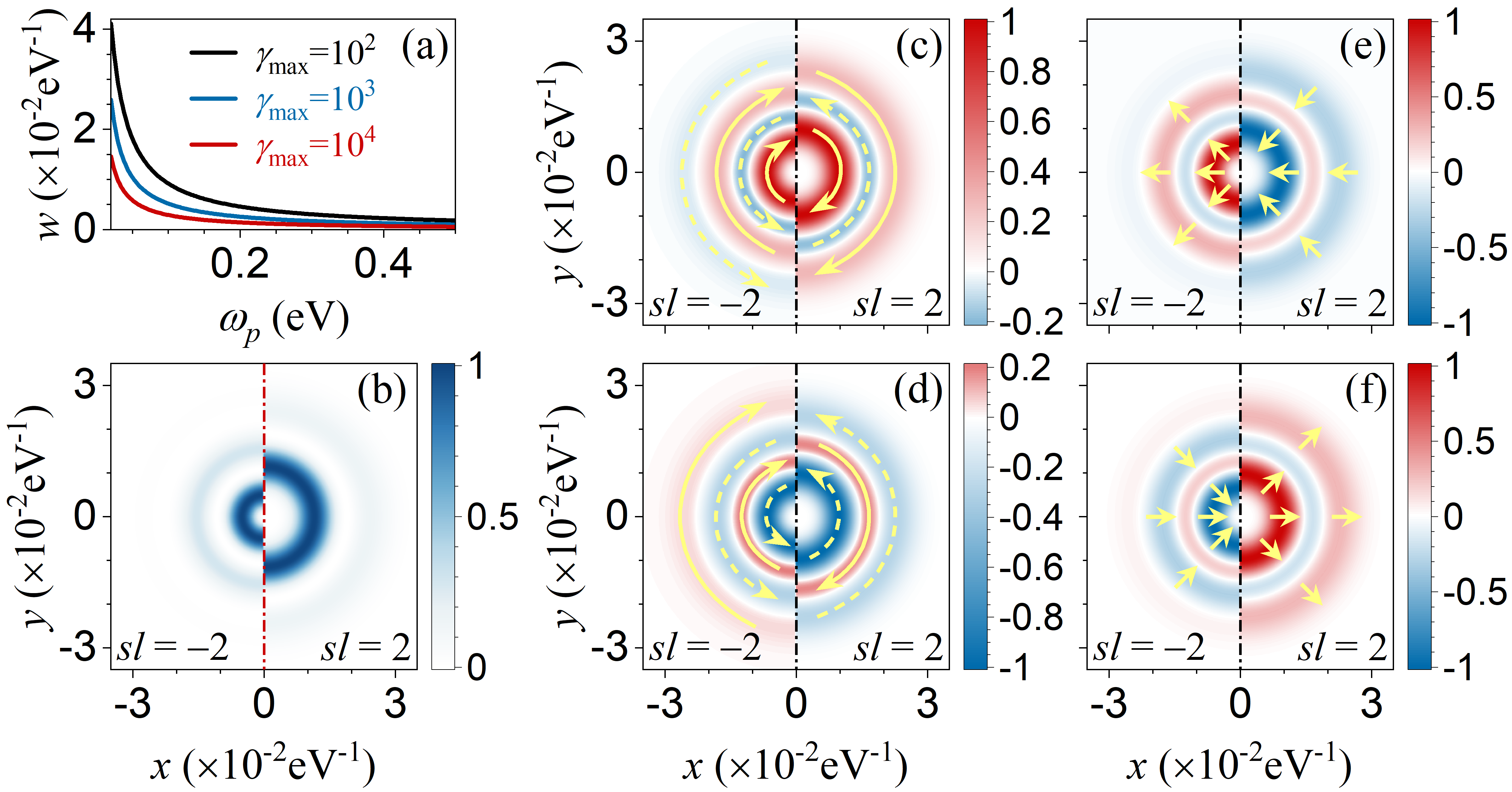}
\caption{Transverse features of eigenstate \eqref{eq:05}. (a) Transverse size $w$ versus plasma frequency for various maximum kinetic energy gain. (b) High-order transverse density for $sl=\pm2$. Vortex current $\bm{j}_{\text{vor}}$ for the OAM number of $l=2$ (c) and $l=-2$ (d). The positive (red) and negative (blue) values represent clockwise and counterclockwise vortex directions, as indicated by yellow arrows. (e) and (f) Radial current $\bm{j}_{\text{rad}}$ for the OAM number of $|l|=2$ at two longitudinal planes separated by $\lambda_{e}/4$ in phase. The red and blue colors represent the outward and inward directions. The radial mode is $n=1$. All the distributions are normalized.}
\label{fig:02}
\end{figure}

Fortunately, the wakefield structure in Fig.~\ref{fig:01} satisfies the crossed-field condition. For convenience, we adopt co-moving variables with the plasma wake: $\tau=t$ and $\xi=z-\beta_{b}t$. The absence of longitudinal magnetic field allows a further simplification: $\Phi_{+}=\Phi_{-}=\Phi$, and Eq.~\eqref{eq:03} is reduced to $[(i\partial_{\mu}-eA_{\mu})^{2}-m_{e}^{2}+ie E_{z}]\Phi(\tau,\bm{r},\xi)=0$. This equation contains a coupling term $\propto r^{2}\partial_{\xi}$, see Appendix A. Assume the transverse size of the electron state to be much smaller than the radius of the plasma cavity, this coupling term is negligible and the transverse and longitudinal modes of $\Phi(\tau,\bm{r},\xi)$ are separable, $\Phi(\tau,\bm{r},\xi)=R^{|l|}_{n}(r)Z(\xi)e^{-iW\tau+il\theta}$, with the OAM number $l$ and conserved energy $W$. The radial mode satisfies
\begin{equation}
\left(\frac{d^{2}}{dr^{2}}+\frac{1}{r}\frac{d}{dr}-\frac{l^{2}}{r^{2}}-\Omega^{4}r^{2}+p_{\perp}^{2}\right)R^{|l|}_{n}(r)=0,
\label{eq:04}
\end{equation}
with $\Omega^{2}=\omega_{p}\sqrt{m_{e}W}/2$. Eq.~\eqref{eq:04} has direct physical meaning: the harmonic-oscillator confinement to electron transverse motion induced by the wakefield focusing force, which immediately gives LG modes $R^{|l|}_{n}(r)=r^{|l|}\exp(-\Omega^{2}r^{2}/2)L^{|l|}_{n}(\Omega^{2}r^{2})$ with quantized transverse momentum $p_{\perp}^{2}=2(2n+|l|+1)\Omega^{2}$. Here $L_{n}^{|l|}$ denotes the associated Laguerre polynomial and $n\geq0$ is the number of radial nodes. $p_{\perp}$ and the effective longitudinal momentum $p_{z}$ satisfy the energy-momentum relation $W^{2}=p_{\perp}^{2}+p_{z}^{2}+m_{e}^{2}$. Defining $m=2n+|l|$, the transverse momentum is degenerate, because all pairs of $(n,l)$ fitting this definition for a given $m$ share the same $p_{\perp,m}$, and its degeneracy is $(m+1)$. This leads to the same degeneracy for electron energy $W_{m}$. The radial LG modes show the central wakefield dressing effect. Compared with two typical free-space vortex states, a free Bessel state has a continuous transverse momentum and is not transversely localized. A free LG state is localized but diffracts unless externally guided. The plasma wake, by contrast, acts as a moving quantum waveguide. Its focusing field fixes the transverse eigenstate to a LG mode with a localized and constant spot, whose size is governed by the plasma frequency and conserved energy, $w\sim1/\Omega=(\omega_{p}\sqrt{m_{e}W}/2)^{-1/2}$. Fig.~\ref{fig:02}(a) shows the transverse size $w$ as functions of plasma frequency and maximum kinetic energy gain. Reducing $\omega_{p}$ increases the bubble radius, thereby enlarging the transverse size. When $\omega_{p}<0.1\,\text{eV}$, $w$ reaches the order of 10\,nm. The size also declines for higher electron energies.

Substituting $\Phi(\tau,\bm{r},\xi)$ into Eq.~\eqref{eq:02} and considering the eigen-spinor, $\sigma_{z}\chi^{s}=s\chi^{s}$, yields the spinor vortex state,
\begin{alignat}{1}
&\Psi_{\alpha}(\tau,\bm{r},\xi)\nonumber\\
=&\,Ne^{-iW\tau+il\theta}\bigg(\begin{array}{c}
R^{|l|}_{n}(r)\Xi^{+}(\xi)\chi^{s}-is e^{is\theta}F^{l}_{n}(r)Z(\xi)\chi^{-s} \\
sR^{|l|}_{n}(r)\Xi^{-}(\xi)\chi^{s}-i e^{is\theta}F^{l}_{n}(r)Z(\xi)\chi^{-s}
\end{array}\bigg).\nonumber\\
\label{eq:05}
\end{alignat}
Here $\alpha$ is an abbreviated notation standing for the collection of all continuous and discrete indices $l$, $n$, $s$ and $W$. The transverse function $F^{l}_{n}(r)=\frac{dR^{|l|}_{n}(r)}{dr}-\frac{sl}{r}R^{|l|}_{n}(r)$ depends on the sign of $(sl)$ and is multiplied by a spin-dependent phase $e^{is\theta}$, thus this term describes the spin-orbit coupling. The detailed forms of $F^{l}_{n}(r)$ and $\Xi^{\pm}(\xi)$ are given in Appendix A. We will show below that this spin–orbit coupling term affects the probability and current density of the electron states. State \eqref{eq:05} is the eigenstate of the electron TAM operator $J_{z}$ with the eigenvalue $l+s/2$. In the paraxial limit, the orbital index can be interpreted as the usual electron OAM number. However, the exact symmetry-governed label is the TAM number.

The probability density from eigenstate \eqref{eq:05} consists of two parts (see Appendix B). The dominant part is a standard LG-type profile such as $\big(R^{|l|}_{n}(r)\big)^{2}$. In addition, there is a higher-order density proportional to $\big(F^{l}_{n}(r)\big)^{2}$, whose structure is determined by the spin-orbit coupling, as shown in Fig.~\ref{fig:02}(b). Reversing the sign of $(sl)$ alters both the radius and the number of rings in the higher-order probability density.

The transverse focusing field induces a radial current vector $\bm{j}_{\text{rad}}$, which, together with the vortex current $\bm{j}_{\text{vor}}$, constitutes the total transverse current $\bm{j}_{\perp}=\bm{j}_{\text{vor}}+\bm{j}_{\text{rad}}$. Both $\bm{j}_{\text{vor}}$ and $\bm{j}_{\text{rad}}$ share the transverse profile $R^{|l|}_{n}(r)F^{l}_{n}(r)$, and their distributions are also critically related to the spin-orbit coupling (see Appendix B). Figs.~\ref{fig:02}(c) and (d) display the transverse distributions of $\bm{j}_{\text{vor}}$ for $l=2$ and $-2$. The sign of $l$ determines the vortex direction, whereas varying the sign of $(sl)$ reshapes the profiles. The radial current $\bm{j}_{\text{rad}}$ has a similar transverse profile, but its direction changes during the longitudinal propagation. Fig.~\ref{fig:02}(e) and (f) show the $\bm{j}_{\text{rad}}$ distributions at two longitudinal planes separated by $\lambda_{e}/4$ in phase, where $\lambda_{e}$ is the electron wavelength. The reversal of $\bm{j}_{\text{rad}}$ between them evidences the longitudinal-mode evolution.

\begin{figure}[b]
\centering
\includegraphics[width=8.6cm]{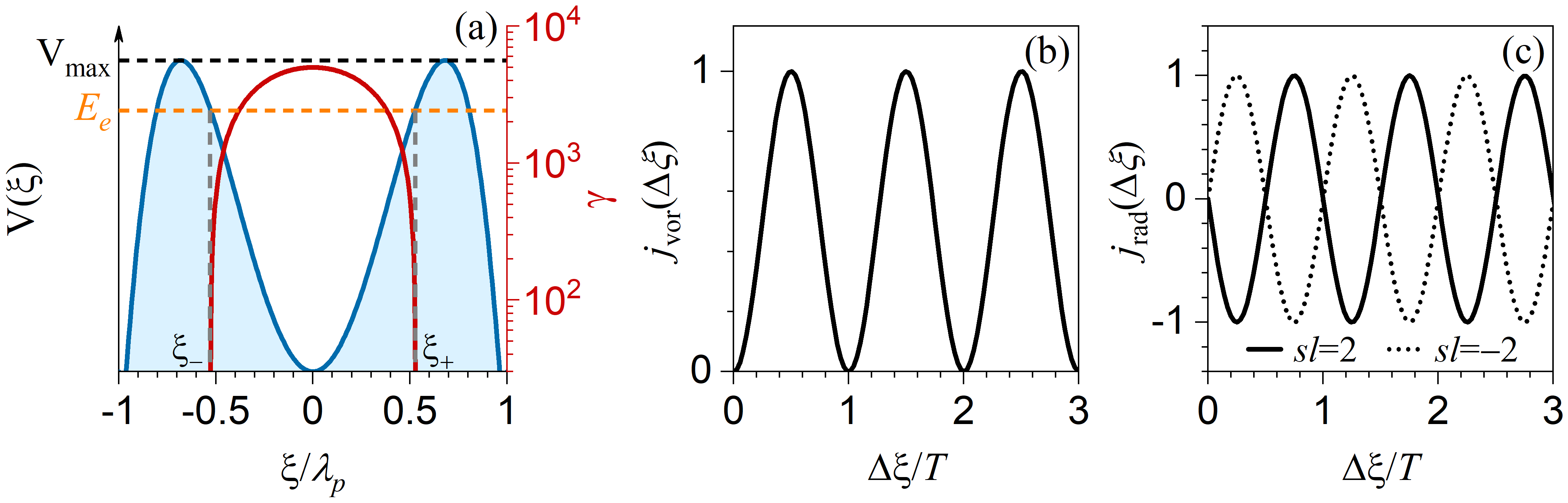}
\caption{Electron acceleration and longitudinal mode evolution. (a) Double-barrier potential $V(\xi)$ in wakefield (blue line) and electron kinetic energy evolution (red line). The electron density in plasma is $n_{0}=1.9\times10^{18}\text{cm}^{-3}$, the relativistic factor of wakefield velocity is $\gamma_{b}=30$. The vortex electron is initially located at $\xi_{-}=0.53\lambda_{p}$ with velocity of $\beta_{0}=\beta_{b}$. Longitudinal evolutions of the normalized current strengths of $\bm{j}_{\text{vor}}$ (b) and $\bm{j}_{\text{rad}}$ (c) in three periods.}
\label{fig:03}
\end{figure}

The longitudinal field governs the acceleration. The longitudinal equation for mode $Z(\xi)$ can be reduced to an effective one-dimensional Schr\"{o}dinger-type equation in a double-barrier potential $V(\xi)$: $(-d^{2}/d\xi^{2}+V(\xi))h(\xi)=E_{e}h(\xi)$, where $h(\xi)$ is the reduced $Z(\xi)$ mode and $E_{e}$ is the electron effective energy, see Appendix A. The blue line in Fig.~\ref{fig:03}(a) shows the double-barrier potential. The electron acceleration (or deceleration) in wakefield can be interpreted as the motion in $V(\xi)$ potential. The barrier height $V_{\text{max}}$ exceeds $E_{e}$, so electrons are confined to oscillate between $\xi_{-}$ and $\xi_{+}$ positions of the barrier. This confinement can modulate the longitudinal modes to produce the wave-node structures. Then the transverse current $\bm{j}_{\perp}$ also oscillates with the longitudinal evolution, with a period of $T=\lambda_{e}/2$. This oscillation alters the strength of the vortex current $\bm{j}_{\text{vor}}$ but leaves its direction intact, see Fig.~\ref{fig:03}(b). In contrast, $\bm{j}_{\text{rad}}$ reverses its direction every half oscillation period (Fig.~\ref{fig:03}(c)), and opposite signs of $(sl)$ introduce a relative phase shift of $T/2$, which are consistent with Fig.~\ref{fig:02}(e) and (f).

The conserved energy $W$ determines the kinetic motion of vortex electrons in PWFA, as shown by the red line in Fig.~\ref{fig:03}(a). The electron gains kinetic energy as it moves toward the center of the wake, where the potential energy is converted into kinetic energy. The maximum kinetic energy gain obeys $\gamma_{\text{max}}\approx2\gamma_{b}^{2}W/m_{e}$ \cite{r30,r42}. The important point for vortex dynamics is that the longitudinal field is $\theta$-independent. Although it changes the kinetic energy and longitudinal momentum of vortex electron, it does not mix the states in different vortex modes. Acceleration thus proceeds within a fixed mode. The transverse field creates and confines the vortex eigenmodes, while the longitudinal field accelerates them without destroying the symmetry-protected vortex phase.

\begin{figure}[b]
\centering
\includegraphics[width=8.6cm]{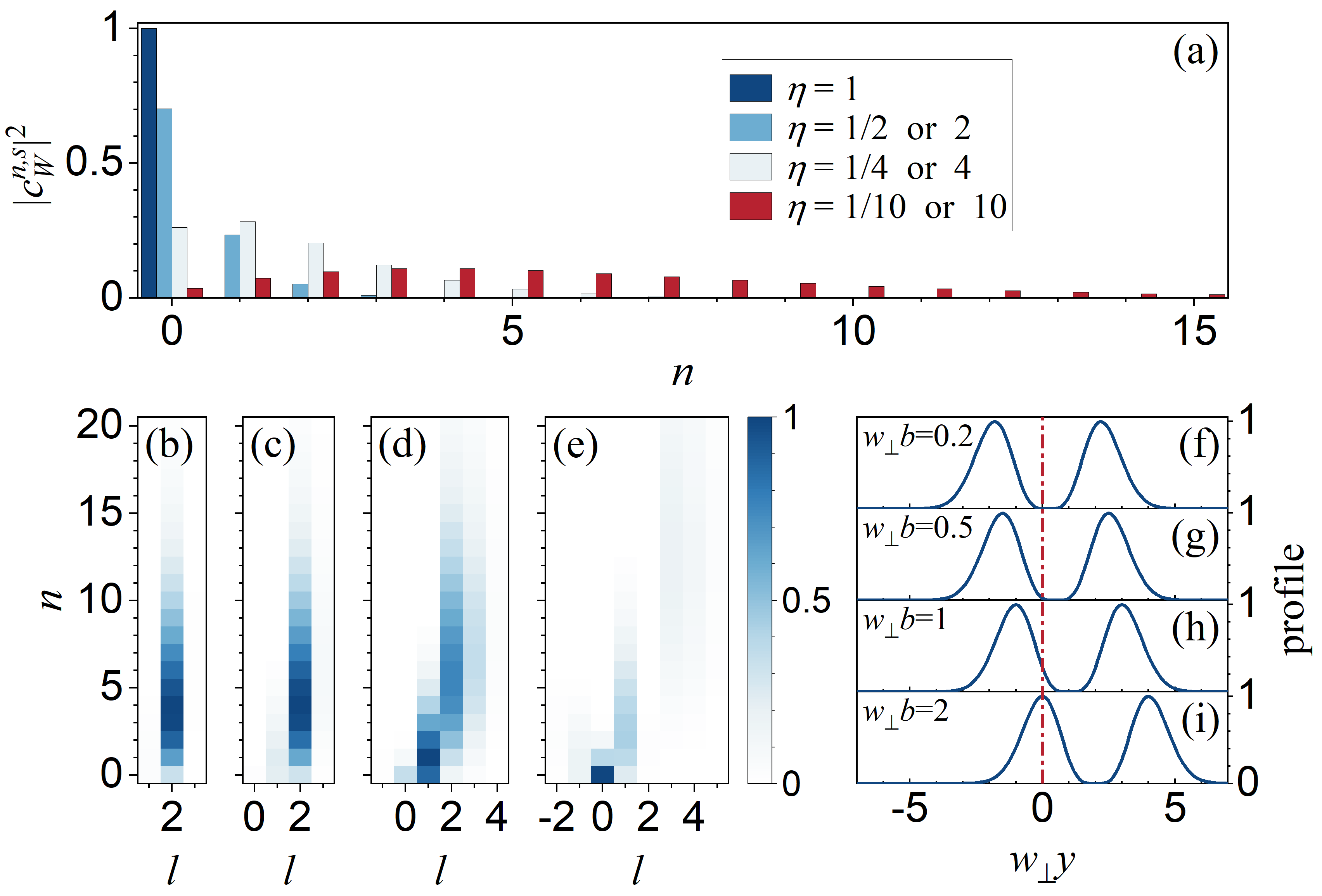}
\caption{Transverse spectra of localized electron wave packet. (a) Radial $n$-spectrum for various transverse size $w_{\perp}$ of the on-axis localized wave packet. (b)–(e) $l$- and $n$-spectra of the off-axis vortex electron wave-packet for different $b$ values, (f)–(i) the corresponding transverse density profiles of electron wave packets as functions of $y$. The initial OAM number of the vortex electron is $l_{0}=2$, its transverse size is $\eta=0.1$.}
\label{fig:04}
\end{figure}

\textit{Vortex electron with localized wave-packet}---The eigenstates \eqref{eq:05} form a natural basis for describing the dynamics of vortex electrons with localized wave packets. Assume that this electron wave-packet is centered at position $z_{0}$ on the propagation axis of the wakefield at time $t=0$, its quantum state is denoted by $\psi^{l_{0}}_{p_{0}}(t,\bm{x})$, where $l_{0}$ and $p_{0}$ are the OAM number and central momentum. This localized wave packet can be expanded over wakefield-dressed eigenstates at time $t=0$: $\psi^{l_{0}}_{p_{0}}(0,\bm{x})=\sum_{\alpha}a_{\alpha}\Psi_{\alpha}(0,\bm{x})$. The expansion coefficients connect with various radial eigenmodes, but involve only the same OAM channel, $a_{\alpha}=c^{n,s}_{W}\delta_{ll_{0}}$, see details in Appendix C. In general, the transverse size $w_{\perp}$ of the wave packet does not match the eigenstate size $\Omega$, so $|c^{n,s}_{W}|^{2}$ yields a radial $n$-spectrum. Fig.~\ref{fig:04}(a) illustrates the dependence of the $n$-spectrum on the transverse size parameter, $\eta=w_{\perp}^{2}/(2\Omega^{2})$. The electron state $\psi^{l_{0}}_{p_{0}}$ coincides with the wakefield-dressed eigenstate of $n=0$ only when their transverse sizes match exactly ($\eta=1$). As the mismatch between $w_{\perp}$ and $\Omega$ increases, the $n$-spectrum broadens. But the OAM-mode is locked because the on-axis wave packet respects the rotational symmetry, and the electron can preserve its vortex phase during acceleration even when its radial profile is a wave packet.

The off-axis wave packet behaves markedly differently, because the rotational symmetry is broken. Assume that the center of the electron wave packet is displaced from the wake axis by a distance $b$ along the y-direction, and the expansion coefficients also contain multiple $l$ components. Taking $l_{0}=2$ as an example, Fig.~\ref{fig:04}(b)–(e) show the $l$- and $n$-spectra for various $b$, the corresponding transverse profiles of the wave packets as a function of $y$ are shown in Fig~\ref{fig:04}(f)–(i). The detailed expansion coefficients are shown in Appendix C. As $b$ increases, the $l$-spectrum broadens, and the dominant OAM component no longer necessarily coincides with $l_{0}$. Different $l$ components have slightly different eigen-energies, $W_{ln}\approx W_{0}+(n+|l|/2+1/2)\omega_{p}\sqrt{m_{e}/W_{0}}$, with $W_{0}=\sqrt{m_{e}^{2}+p_{z}^{2}}$, and therefore accumulate different dynamical phases, $e^{-iWt}$, during propagation. After propagating over a time-scale of plasma wavelength, these phase differences become non-negligible and induce dispersion among the different modes. The $n$-dispersion mainly deforms the transverse profile of the wave packet, whereas the $l$-dispersion directly degrades the vortex phase. To suppress such dispersion and the $l$-spectrum broadening, a sufficiently small off‑axis offset is required. Under the parameters in Fig.~\ref{fig:04}(b)–(e), the OAM number remains nearly single‑valued when the offset satisfies $w_{\perp}b<0.5$, and the wakefield does not destroy the vortex phase and TAM conservation in these cases. 

The $n$- and $l$-dispersion may also induce transverse oscillation of the entire electron wave packet in the wakefield, producing betatron radiation. Recent studies have shown that the OAM transfer for electrons is insignificant when radiation reaction is negligible \cite{r26,r43}, which is the case in regular regimes of wakefield acceleration.

\textit{Non-ideal wakefield}---Finally, realistic plasma wakefields are not perfectly axisymmetric and may deviate slightly from the crossed-field condition. Such effects can be treated as perturbations to the ideal Hamiltonian $H(r,\theta,\xi)=H_{0}(r,\xi)+\tilde{H}(r,\theta,\xi)$, where $H_{0}$ is the Hamiltonian in ideal wakefield, $\tilde{H}$ accounts for the Hamiltonian of asymmetric or non-crossed-field perturbations. To the first order, the effect of  $\tilde{H}$ is governed by the matrix elements of $\langle l',n'|\tilde{H}|l,n\rangle=\int d^{3}x\Psi_{l'n'}^{\dag}\tilde{H}\Psi_{ln}$, where $\Psi_{l'n'}$ is the eigenstate in non-degenerate space and $\Psi_{ln}$ is the one in degenerate space (see details in Appendix D). Perturbations of the nonideal wakefield may modify the radial profiles and mix the angular-momentum modes of the eigenstates. Expanding the perturbed Hamiltonian into angular harmonics, $\tilde{H}(r,\theta,\xi)=\sum_{\varepsilon}\tilde{H}_{\varepsilon}(r,\xi)e^{i\varepsilon\theta}$, the azimuthal integration yields the selection rule via a $\delta$-function, $\delta_{\varepsilon,l'-l}$. The azimuthal perturbation with harmonic number $\varepsilon$ shifts the electron angular momentum by $\varepsilon$ units according to the above selection rule. Thus, the nonaxisymmetric harmonics can couple different angular-momentum channels to broaden the angular-momentum spectrum. By contrast, For a perturbation with a single vortex phase, $\tilde{H}(r,\theta,\xi)=\tilde{H}(r,\xi)e^{i\varepsilon_{0}\theta}$, the angular-momentum mixing is restricted to the corresponding channel, and the number of coupled modes is bounded by the degeneracy of the unperturbed transverse level. These selection rules provide a direct criterion for assessing vortex-state degradation in nonideal wakefields.

\textit{Conclusions}---We have shown that an axisymmetric plasma wakefield can support total-angular-momentum eigenstates of relativistic vortex electrons. These wakefield-dressed spinor states provide a natural basis for describing the acceleration of arbitrary vortex-electron wave packets in plasma wakefield structures. The transverse focusing field acts as a moving quantum waveguide, producing Laguerre-Gaussian-like modes with discrete transverse momenta, while the longitudinal electric field accelerates the electron without breaking the symmetry-protected total angular momentum. We have further analyzed the radial- and angular-momentum mode broadening induced by localized wave packets, off-axis injection, and nonideal wakefield perturbations, and identified the physical conditions required to preserve the vortex structure. The framework developed here applies more generally to axisymmetric crossed-field configurations and goes beyond nonrelativistic, scalar, and quasiclassical descriptions. Our results extend plasma wakefield acceleration from classical beam dynamics to the control of microscopic quantum states of high-energy electrons.

\bigskip

\textit{Acknowledgments}---This work was supported by the National Science Foundation of China (Grant No. 12388102), the Strategic Priority Research Program of the Chinese Academy of Sciences (Grant No. XDB0890303), and the CAS Project for Young Scientists in Basic Research (Grant No.YSBR060).

% The \nocite command causes all entries in a bibliography to be printed out
% whether or not they are actually referenced in the text. This is appropriate
% for the sample file to show the different styles of references, but authors
% most likely will not want to use it.
\nocite{*}

\bibliography{QVS_acceleration}% Produces the bibliography via BibTeX.

\clearpageandresetcolumns

\appendix

\setcounter{equation}{0}
\renewcommand{\theequation}{A\arabic{equation}}

\section*{End Matter}

\textit{Appendix A: Solving scalar state $\Phi(x)$}---In the electromagnetic potential of the plasma wake in spherical cavity model \cite{r39}, the wakefield has no longitudinal magnetic field, Eq.~\eqref{eq:02} is simplified, in the co-moving variables, to
\begin{alignat}{1}
&\bigg[-\frac{\partial^{2}}{\partial\tau^{2}}+2\beta_{b}\frac{\partial^{2}}{\partial\tau\partial\xi}+\frac{1}{\gamma_{b}^{2}}\frac{\partial^{2}}{\partial\xi^{2}}+\nabla_{\perp}^{2}-\frac{im_{e}\omega_{p}^{2}}{4}\left(r^{2}+\xi^{2}\right)\nonumber\\
&\times\bigg(\frac{\partial}{\partial\tau}-(1+\beta_{b})\frac{\partial}{\partial\xi}\bigg)-m_{e}^{2}\bigg]\Phi(x)=0.
\label{eq:A1}
\end{alignat}
The time $\tau$ in Eq.~\eqref{eq:A1} is separable, but there is a coupling between the transverse and longitudinal coordinates: $r^{2}\partial_{\xi}$. Considering that the transverse size of the electron state is much smaller than the radius of the plasma wake, $r\ll R_{0}$, the coupling term $r^{2}\partial_{\xi}$ is ignorable. Then, the transverse and longitudinal coordinates become separable, and the solution takes the form, $\Phi(\tau,\bm{r},\xi)=R(r,\theta)Z(\xi)e^{-iW\tau}$. The transverse and longitudinal modes satisfy the equations,
\begin{equation}
\left(\nabla_{\perp}^{2}-\Omega^{4}r^{2}+p_{\perp}^{2}\right)R(r,\theta)=0
\label{eq:A2}
\end{equation}
and
\begin{alignat}{1}
&\bigg[\frac{d^{2}}{d\xi^{2}}+i\gamma_{b}^{2}\bigg(\frac{(1+\beta_{b})\Omega^{4}}{W}\xi^{2}-2\beta_{b}W\bigg)\frac{d}{d\xi}\nonumber\\
&-\Omega^{4}\gamma_{b}^{2}\xi^{2}+\gamma_{b}^{2}p_{z}^{2}\bigg]Z(\xi)=0.
\label{eq:A3}
\end{alignat}
The solution to Eq.~\eqref{eq:A2} is $R(r,\theta)=R^{|l|}_{n}(r)e^{il\theta}$, with $R^{|l|}_{n}(r)=r^{|l|}\exp(-\Omega^{2}r^{2}/2)L^{|l|}_{n}(\Omega^{2}r^{2})$ being a standard LG mode. Inserting the solution $\Phi(\tau,\bm{r},\xi)$ into Eq.~\eqref{eq:03} and consider the free spinor $\chi^{s}$ as the eigenstate of $\sigma_{z}$, the spinor solution, Eq.~\eqref{eq:05}, is obtained. The transverse function $F^{l}_{n}(r)$ is determined by the sign of $(sl)$,
\begin{alignat}{1}
&F^{l}_{n}(r)=\frac{dR^{|l|}_{n}(r)}{dr}-\frac{sl}{r}R^{|l|}_{n}(r)\nonumber\\
=&\left\{\begin{array}{cc}
-\Omega^{2}\left(R^{|l|+1}_{n}(r)+R^{|l|+1}_{n-1}(r)\right), & \text{for}~~sl>0; \\
(n+|l|)R^{|l|-1}_{n}(r)+(n+1)R^{|l|-1}_{n+1}(r), & \text{for}~~sl<0.
\end{array}\right.\nonumber\\
\label{eq:A4}
\end{alignat}
The longitudinal function $\Xi^{\pm}(\xi)=(W\pm m_{e})Z(\xi)-i(1+\beta_{b})Z'(\xi)$, with the derivative $Z'(\xi)=dZ(\xi)/d\xi$.

Assume the longitudinal mode $Z(\xi)=\exp\left[-i\gamma_{b}^{2}(m_{e}(1+\beta_{b})\omega_{p}^{2}\xi^{3}/24-\beta_{b}W\xi)\right]h(\xi)$, we derive a one-dimensional Schrödinger-type equation for $h(\xi)$: $(-d^{2}/d\xi^{2}+V(\xi))h(\xi)=E_{e}h(\xi)$, where the effective energy $E_{e}=\gamma_{b}^{2}(\beta_{b}^{2}\gamma_{b}^{2}W^{2}+K_{z})$, and the potential,
\begin{equation}
V(\xi)=-\frac{\Omega^{4}}{(1-\beta_{b})}\bigg(\frac{\Omega^{4}}{4W^{2}(1-\beta_{b})}\xi^{4}-\gamma_{b}^{2}\xi^{2}-\frac{i\xi}{W}\bigg).
\label{eq:A5}
\end{equation}
Eq.~\eqref{eq:A5} contains a non-Hermitian term $im_{e}\omega_{p}^{2}\xi/4(1-\beta_{b})$, which becomes significant only as $\xi\to0$, thus can be neglected. The remaining potential represents a typical double-barrier structure. At the position $\xi=\xi_{0}=\pm\frac{\lambda_{p}}{\pi}\sqrt{\frac{2W}{m_{e}(1+\beta_{b})}}$, the potential barrier reaches its maximum value $V_{\text{max}}(\xi_{0})=\gamma_{b}^{4}W^{2}$. Since $V_{\text{max}}>E_{e}$, the electron cannot overcome the barrier and only oscillate between $\xi_{-}$ and $\xi_{+}$ in the classical picture, here $\xi_{\pm}=\pm\frac{\lambda_{p}}{\pi}\Big[\frac{2(\gamma_{b}W-\sqrt{m_{e}^{2}+p_{\perp}^{2}})}{m_{e}\gamma_{b}(1+\beta_{b})}\Big]^{1/2}$. In the quantum picture, it is possible for an electron to traverse the barrier via the tunneling effect. However, the width of the portion of the barrier height exceeding $E_{e}$ is on the order of $1/(\gamma_{b}\omega_{p})$, significantly larger than the electron wavelength. Therefore, the tunneling probability is extremely low and can be neglected, and electrons must be injected in the region of $\xi_{-}<\xi<0$ to be effectively accelerated. If $|\xi_{\pm}|\leq R_{0}$, the initial electron velocity at $\xi_{-}$ must satisfy $\beta_{0}=\beta_{b}$.

We can obtain an approximate solution for mode $Z(\xi)$ in the region of $\omega_{p}^{2}\xi^{2}\ll1$, where Eq.~\eqref{eq:A3} is reduced to
\begin{equation}
\bigg(\frac{d^{2}}{d\xi^{2}}-2i\beta_{b}\gamma_{b}^{2}W\frac{d}{d\xi}-\Omega^{4}\gamma_{b}^{2}\xi^{2}+\gamma_{b}^{2}p_{z}^{2}\bigg)Z(\xi)=0.
\label{eq:A6}
\end{equation}
Eq.~\eqref{eq:A6} has an analytical solution, $Z(\xi)=D_{\nu}(\Omega\sqrt{2\gamma_{b}}\xi)e^{i\beta_{b}\gamma_{b}^{2}W\xi}$, where $D_{\nu}(v)$ is the parabolic cylinder function, and parameter $\nu=E_{e}/(2\gamma_{b}\Omega^{2})-1/2$.

\setcounter{equation}{0}
\renewcommand{\theequation}{B\arabic{equation}}

\textit{Appendix B: Probability and current densities of TAM eigenstate}---Using the TAM eigenstate in wakefield, Eq.~\eqref{eq:05}, the four-dimensional current density $j^{\mu}(x)=\bar{\Psi}^{l,s}_{n}(x)\gamma^{\mu}\Psi^{l,s}_{n}(x)$ is derived. The probability density is given by its 0-component, with the result,
\begin{alignat}{1}
\rho^{l,s}_{n}(x)\sim&\,\big(R^{|l|}_{n}(r)\big)^{2}\left[(W^{2}+M^{2})|Z(\xi)|^{2}\right.\nonumber\\
&\left.+2(1+\beta_{b})W\text{Im}(Z'(\xi)Z^{*}(\xi))\right.\nonumber\\
&\left.+(1+\beta_{b})^{2}|Z'(\xi)|^{2}\right]+\big(F^{l}_{n}(r)\big)^{2}|Z(\xi)|^{2}.
\label{eq:B1}
\end{alignat}
The spatial component of $j^{\mu}(x)$ gives the current density vector. Since the wakefield has transverse focusing field, the transverse current density contains the vortex current and radial current components, $\bm{j}_{\perp}(t,\bm{x})=\bm{j}_{\text{vor}}(t,\bm{x})+\bm{j}_{\text{rad}}(t,\bm{x})$, where the vortex current is given by
\begin{alignat}{1}
\bm{j}_{\text{vor}}(t,\bm{x})\sim&\,sR^{|l|}_{n}(r)F^{l}_{n}(r)\left[(1+\beta_{b})\text{Im}(Z'(\xi)Z^{*}(\xi))\right.\nonumber\\
&\left.+W|Z(\xi)|^{2}\right](\sin\theta\bm{e}_{x}-\cos\theta\bm{e}_{y}),
\label{eq:B2}
\end{alignat}
and the radial current takes the form
\begin{alignat}{1}
\bm{j}_{\text{rad}}(t,\bm{x})\sim&\,(1+\beta_{b})R^{|l|}_{n}(r)F^{l}_{n}(r)\text{Re}(Z'(\xi)Z^{*}(\xi))\nonumber\\
&\times(\cos\theta\bm{e}_{x}+\sin\theta\bm{e}_{y}).
\label{eq:B3}
\end{alignat}

\setcounter{equation}{0}
\renewcommand{\theequation}{C\arabic{equation}}

\textit{Appendix C: Transverse mode spectrum of localized vortex wave packet}---Suppose a vortex electron with localized wave packet is located at the position $z_{0}$ on the propagation axis of the wakefield at time $t=0$, its wave-function can be constructed by superimposing plane-wave states with specific vortex phase, $\psi^{l_{0}}_{p_{0}}(x)\sim\int d^{3}pe^{-iE_{p}t+i\bm{p}\cdot\bm{x}}u(\bm{p})\tilde\psi^{l_{0}}_{p_{0}}(\bm{p})$, where $\tilde\psi^{l_{0}}_{p_{0}}(\bm{p})=|\bm{p}_{\perp}|^{|l_{0}|}\exp[-\bm{p}_{\perp}^{2}/w_{\perp}^{2}-(p_{z}-p_{0})^{2}/w_{z}^{2}-ip_{z}z_{0}+il_{0}\theta_{p}]$ is the momentum spectrum, $l_{0}$ and $p_{0}$ are the OAM number and central momentum of the wave packet, $w_{\perp/z}\ll p_{0}$ are the transverse/longitudinal wave packet width in momentum space, $u(\bm{p})$ is the free spinor wave-function. In the relativistic case, $m_{e}\ll E_{p}$, the lowest-order wave-function is obtained,
\begin{alignat}{1}
\psi^{l_{0}}_{p_{0}}(t,\bm{x})\approx&\,Cr^{|l_{0}|}\exp\left[-\frac{w_{\perp}^{2}r^{2}}{4}-\frac{w_{z}^{2}(z-z_{0}-\beta_{z}t)^{2}}{4}\right]\nonumber\\
&\times\exp[-iE_{0}t+il_{0}\theta+ip_{0}(z-z_{0})]\Big(\begin{array}{c}
u \\
\sigma_{z}u
\end{array}\Big),\nonumber\\
\label{eq:C1}
\end{alignat}
where $C$ is the normalization coefficient and $u$ is the two-component spinor wave-function of the free electron. For simplicity, we use the following approximations. The transverse momentum of eigenstate \eqref{eq:05} is much smaller than electron mass; under the typical wakefield parameters, we have $WZ(\xi)\sim m_{e}Z(\xi)\ll dZ(\xi)/d\xi$, allowing us to retain only the term proportional to $dZ(\xi)/d\xi$ in Eq.~\eqref{eq:05}. These conditions reduce the eigenstate \eqref{eq:05} to a simpler form, $\Psi_{\alpha}(t,\bm{x})\approx\frac{N}{\sqrt{2}}R^{|l|}_{n}(r)\frac{dZ(\xi)}{d\xi}e^{-iWt+il\theta}\left(\begin{array}{c}
\chi^{s} \\
s\chi^{s}
\end{array}\right)$. If we expand the localized electron wave-packet over this eigenmode at $t=0$, the expansion coefficient is $a_{\alpha}=\int d^{3}x\Psi^{\dag}_{\alpha}(0,\bm{x})\psi^{l_{0}}_{p_{0}}(0,\bm{x})=c^{n,s}_{W}\delta_{ll_{0}}$, where $c^{n,s}_{W}$ is given by,
\begin{alignat}{1}
&c^{n,s}_{W}(\eta)\nonumber\\
=&\frac{2^{|l_{0}|+1}}{(2\pi)^{1/4}}\sqrt{\frac{(n+|l_{0})!}{n!|l_{0}|!}}\frac{\eta^{(|l_{0}|+1)/2}(\eta-1)^{n}}{(\eta+1)^{n+|l_{0}|+1}}\rho_{z}(W)(\chi^{s\dag}u),\nonumber\\
\label{eq:C2}
\end{alignat}
with
\begin{alignat}{1}
&\rho_{z}(W)\nonumber\\
=&\,\sqrt{w_{z}}\int dz\frac{dZ^{*}(z)}{dz}\exp\left[ip_{0}(z-z_{0})-\frac{w_{z}^{2}(z-z_{0})^{2}}{4}\right],\nonumber\\
\label{eq:C3}
\end{alignat}
and $\eta=w_{\perp}^{2}/(2\Omega^{2})$. $|c^{n,s}_{W}|^{2}$ gives the radial $n$-spectrum.

If the electron wave-packet is off-axis with an offset distance $b$ along the y-axis, it takes the form
\begin{alignat}{1}
&\psi^{l_{0}}_{p_{0}}(0,\bm{x})\nonumber\\
\approx&\,C\left[x^{2}+(y-b)^{2}\right]^{|l_{0}|/2}\exp\left[-\frac{w_{\perp}^{2}(x^{2}+(y-b)^{2})}{4}\right]\nonumber\\
&\times\exp\left[il_{0}\left(\theta-\arctan\left(\frac{b\cos\theta}{r-b\sin\theta}\right)\right)\right]\nonumber\\
&\times\exp\left[-\frac{w_{z}^{2}(z-z_{0})^{2}}{4}+ip_{0}(z-z_{0})\right]\Big(\begin{array}{c}
u \\
\sigma_{z}u
\end{array}\Big).
\label{eq:C4}
\end{alignat}
Expanding this state over the eigenstates, the coefficients determine the transverse $l$- and $n$-spectra. Taking $l_{0}=2$ as an example, the coefficient is derived as $a_{\alpha}=(2\pi)^{-1/4}\rho_{\perp}(l,n;\eta)\rho_{z}(W)(\chi^{s\dag}u)$, with
\begin{alignat}{1}
&\rho_{\perp}(l,n;\eta)\nonumber\\
=&\,-\sqrt{\frac{n!}{(n+|l|)!}}\frac{i^{-l}}{2^{(|l|+3)/2}\eta^{(|l|+1)/2}}\exp\left(-\frac{w_{\perp}^{2}b^{2}}{4}\right)\nonumber\\
&\times\int_{0}^{\infty}d\zeta\zeta^{|l|+1}\left[\zeta^{2}I_{l-2}\left(\frac{w_{\perp}b\zeta}{2}\right)\right.\nonumber\\
&\left.-2w_{\perp}b\zeta I_{l-1}\left(\frac{w_{\perp}b\zeta}{2}\right)+w_{\perp}^{2}b^{2}I_{l}\left(\frac{w_{\perp}b\zeta}{2}\right)\right]\nonumber\\
&\times\exp\left[-\frac{\zeta^{2}}{4}\left(1+\frac{1}{\eta}\right)\right]L^{|l|}_{n}\left(\frac{\zeta^{2}}{2\eta}\right),
\label{eq:C5}
\end{alignat}
where $I_{l}(v)$ is the modified Bessel function.

\setcounter{equation}{0}
\renewcommand{\theequation}{D\arabic{equation}}

\textit{Appendix D: Perturbation theory in non-ideal wakefield}---Eq.~\eqref{eq:05} is the eigenstate of the unperturbed Hamiltonian $H_{0}$, its eigenenergy gives the zeroth-order energy, $W^{(0)}_{m}=W_{m}$, which is $(m+1)$-fold degenerate. We can use the degenerate states to construct a superposition state, $\Psi^{(0)}_{m}=\sum\limits_{2n+|l|=m}f_{ln}\Psi_{\alpha}$. Substituting it into Dirac equation yields a determinantal equation for the first-order energy correction $W^{(1)}$, $\text{det}|\langle l',n'|\tilde{H}|l,n\rangle-W^{(1)}\delta_{l'l}\delta_{n'n}|=0$. This equation has $(m+1)$ solutions: $W^{(1)}_{k}$ with $k=1,2,\dots,m+1$, corresponding to $(m+1)$ first-order split energies, each associated with a set of coefficients $f_{ln}^{k}$ and a corresponding split wave-function. The first-order corrected wave-function is given by
\begin{equation}
\Psi^{(1)}_{m,k}=\sum_{\substack{2n'+|l'|\neq m \\ 2n+|l|=m}}\frac{f_{ln}^{k}\langle l',n'|\tilde{H}|l,n\rangle}{W^{(0)}_{m}-W^{(0)}_{l'n'}}\Psi_{\alpha'},
\label{eq:D1}
\end{equation}
which is a linear combination of unperturbed eigenstates. Consequently, the perturbed wakefields lift the degeneracy and mixes different OAM modes, leading to spectral broadening.
If we expand the perturbed Hamiltonian into angular harmonics, $\tilde{H}(r,\theta,\xi)=\sum\limits_{\varepsilon}\tilde{H}_{\varepsilon}(r,\xi)e^{i\varepsilon\theta}$, each harmonic generates a delta-function for the matrix elements: $\langle l',n'|\tilde{H}_{\varepsilon}(r,\xi)e^{i\varepsilon\theta}|l,n\rangle\propto(\tilde{H}_{\varepsilon})_{l'n',ln}\delta_{\varepsilon,l'-l}$. Then, the first-order corrected wave-function reads,
\begin{equation}
\Psi^{(1)}_{m,k}\sim\sum_{\substack{2n'+|l'|\neq m \\ 2n+|l|=m}}\frac{f_{ln}^{k}(\tilde{H}_{l'-l})_{l'n';ln}}{W^{(0)}_{m}-W^{(0)}_{l'n'}}\Psi_{\alpha'}.
\label{eq:D2}
\end{equation}
Here the superposition of eigenstates is carried out in the entire non-degenerate space; therefore, the first-order wavefunction exhibits a broad OAM spectrum. If the perturbation has a definite vortex phase, $\tilde{H}(r,\theta,\xi)=\tilde{H}(r,\xi)e^{i\varepsilon_{0}\theta}$, the first-order wave-function is simplified to,
\begin{equation}
\Psi^{(1)}_{m,k}\sim\sum_{\substack{2n'+|l+\varepsilon_{0}|\neq m \\ 2n+|l|=m}}\frac{f_{ln}^{k}\tilde{H}_{l+\varepsilon_{0},n';ln}}{W^{(0)}_{m}-W^{(0)}_{l+\varepsilon_{0},n'}}\Psi_{l+\varepsilon_{0},n'},
\label{eq:D3}
\end{equation}
where the summation over the OAM numbers in eigenstates is constrained within the degenerate space, thereby compressing the OAM-spectral width.

\end{document}